\documentclass[aps,prd,twocolumn,nofootinbib]{revtex4-1}
\usepackage{epsfig}
\usepackage[colorlinks,linkcolor=blue,anchorcolor=blue,citecolor=blue,urlcolor=blue,breaklinks=true]{hyperref}
\usepackage{graphics}
\usepackage{slashed}
\usepackage{color}
\usepackage{amsfonts}
\usepackage{amsmath}
\usepackage{extarrows}
\usepackage{amssymb}
\usepackage{epstopdf}
\usepackage{float}
\usepackage{multirow}

\begin{document}

\title{The study of exotic state $Z_c^{\pm}(3900)$ decaying to $J/\psi\pi^{\pm}$ \\in the $pp$ collisions at $\sqrt{s}$ = 1.96, 7, and 13~TeV}

\author{Zhen Zhang$^{1}$}\email{Email:2646691260@cug.edu.cn}
\author{Liang Zheng$^{1}$}
\author{Gang Chen$^{1}$}\email{Email:chengang1@cug.edu.cn(Corresponding Author)}
\author{Hong-Ge Xu$^{1}$}
\author{Dai-Mei Zhou$^{2}$}
\author{Yu-Liang Yan$^3$}
\author{Ben-Hao Sa$^3$}
%

\address{$^{1}$ School of Mathematics and Physics, China University of Geosciences, Lumo Road 388, 430074 Wuhan, China.\\
${^2}$Institute of Particle Physics, Huazhong Normal University, Wuhan 430082, China\\
${^3}$China Institute of Atomic Energy, P.O. Box 275(10), Beijing 102413, China}
%

\begin{abstract}
\bigskip
A dynamically constrained phase-space coalescence model and PACIAE model are used to predict the exotic resonant state $Z_c^{\pm}(3900)$ yield in $pp$ collisions at $\sqrt{s} = 1.96, 7$ and 13 TeV, respectively, which are estimated to be around  $10^{-6}$ to $10^{-5}$ based on the $J/\psi\pi^{\pm}$ bound state in the decay chain of $b$ hadrons. The energy dependence of the transverse momentum distributions and rapidity distributions with $|y|<6$ and $p_{T}<10$ GeV/c are also calculated for ${Z_c^{+}(3900)}$  and ${Z_c^{-}(3900)}$. The production of ${Z_c^{+}(3900)}$  and its anti-particle ${Z_c^{-}(3900)}$  are found to be quite similar to each other.

\bigskip
\end{abstract}
\maketitle

\section{Introduction}
Particle physicists believe that quarks are the building block for the matter in our viable universe. Due to the color confinement of strong interaction, quarks are bound into the color neutral hadrons with different configurations. Mesons consisting of quark and antiquark pairs and baryons made of three quarks are the most common hadrons observed in high energy collision experiments. However, other unconventional configurations with more quarks or gluons are also allowed in the quark model framework, for example, the multi-quark states~\cite {r1,r2,r12} composed of 4 or more quarks, the hadronic molecules~\cite {r3,r4,r5}  bound together by hadrons, the hybrid states~\cite {r6,r16} composed of quarks and gluons, and the glueballs composed of gluons. These unconventional hadrons are usually called the exotic state hadrons.

\par In 2013, the BES\uppercase\expandafter{\romannumeral3} Collaboration~\cite {r7,r11,r13} analyzed the invariant mass spectrum of $\pi^{\pm}J/\psi$ in the process $e^{+}e^{-}\rightarrow\pi^{+}\pi^{-}J/\psi$ at $\sqrt{s} = 4.26$ GeV, and found there was a resonance structure around 3.9 GeV/$c^{2}$, whose decay width is $46 \pm 10\pm 20$ MeV.  BES\uppercase\expandafter{\romannumeral3} named it $Z_c^{\pm}(3900)$~\cite {r14,r15,r17}, and in the later experiment, its spin and parity were found to be $J^{P}=1^{+}$~\cite {r8}. This observation has also been confirmed by Belle and CLEO-c experiments~\cite {r9,r17}.

It is speculated based on the experimental data that $Z_c^{\pm}(3900)$ consists of at least four quarks: $c\overline{c}u\overline{d}$ or $c\overline{c}\overline{u}d$, and can either be a Tetraquark state~\cite {r7,r9} or a weakly bounded molecular state considering that the mass of $Z_c^{\pm}(3900)$ is slightly higher than the open-charm $D^{\ast}\overline{D}$ threshold~\cite {r7}.

\par The $D0$ experiment~\cite {r18,r23,r24} speculates that $Z_c^{\pm}(3900)$ might be produced by these two processes $b$ hadron $\rightarrow Y(4260)+h$ and $Y(4260)\rightarrow Z_c^{\pm}(3900)\pi^{\pm}$, where $h$ is any particle other than $Y(4260)$ that produced by the decay of $b$-flavored hadrons~\cite {r19}. By studying the data collected in the $p\bar{p}$ collision, the $D0$ experimental group found the resonant state $Z_c^{\pm}(3900)$ in the invariant mass spectrum of $\pi^{\pm}J/\psi$, and confirmed~\cite {r20} the correlation between the resonant state and $J/\psi\pi^{+}\pi^{-}$ with the invariant mass within the range of $4.2-4.7$ GeV, in which $J/\psi$~\cite {r21,r22} was derived from the $b$-flavored hadron decay. This shows that there is an intermediate state in the decay of $b$-flavored hadron and then decays into $Z_c^{\pm}(3900)$. These observations indicate that further studies on the property of the exotic hadrons would help to understand the formation of exotic hadron states and the nature of the strong force.

In this paper, we treat the $Z_c^{\pm}(3900)$ as a molecular state consisting of $J/\psi \pi^{\pm}$ and present a systematic study on its production in $pp$ collisions based on a Monte Carlo simulation approach. First, event samples of multiparticle final states $J/\psi$, $\pi^{+}$ and $\pi^{-}$ are generated in $pp$ collisions at $\sqrt s=1.96, 7$ and 13~TeV using the parton and hadron cascade model (PACIAE)~\cite {r25}. Then the bound states $J/\psi\pi^{\pm}$ are formed using a dynamically constrained phase space coalescence model (DCPC)~\cite {r31} to study $Z_c^{\pm}(3900)$.

\section{The PACIAE and DCPC model}
The PACIAE model~\cite {r25}, also known as the parton and hadron cascade model, is based on PYTHIA model~\cite {r26}. The PACIAE model is a theoretical model to describe various high-energy collisions, which is used to simulate the $pp$ collision in this paper. The PACIAE model divides high-energy collisions into four stages: parton initiation, parton rescattering, hadronization, and hadron rescattering.

In the first stage, the initial parton conditions are obtained by breaking down the PYTHIA strings created in hard scattering and parton shower into quarks and gluons. After that, further parton-parton rescatterings can happen in the quark-gluon system to model the evolution of the deconfined quark matter state. A $K$ factor is allowed to account for higher order effects in hard scattering and parton-parton rescatterings. After all parton rescatterings, the final state partons are converted to hadrons via the Lund string fragmentation model~\cite {r26} or the coalescence model~\cite{r25}. The last stage is hadron rescattering, and the method of two-body collision~\cite {r28} is used to rescatter the hadronic matter until hadronic freeze-out. More details can be found in Ref\cite {r25}.

In this paper, the yield of nuclei or bound states is calculated in two steps. First, the hadrons are calculated by the PACIAE model. Then, the bound states or exotic states are calculated by the DCPC model, which has been successfully applied to calculate the yield of particles in Pb-Pb~\cite {r29}, Au-Au~\cite {r30,r31,r32,r33} and $pp$ collisions~\cite {r34}.

According to quantum statistical mechanics~\cite {r35}, the yield of particles can be estimated by the uncertainty principle. The yield of single particle can be calculated with the following integral:
\begin{equation}\label{eq1}
  Y_{1}=\int_{E_a\leq H\leq E_b}\frac{d\vec{q}d\vec{p}}{h^{3}}.
\end{equation}
where $E_a, E_b$, and $H$ denote energy threshold and the energy function of the particle, respectively. The variables $\vec q$ and $\vec p$ are the coordinates and momentum of the particle in the center-of-mass frame system at the moment after hadronic completion. Similarly, the yield of a cluster consisting of N particles can be calculated as following:
\begin{equation}\label{eq2}
  Y_{N}=\int\cdots\int_{E_a\leq H\leq E_b}\frac{d\vec{q}_{1}d\vec{p}_{1}\cdots d\vec{q}_{N}d\vec{p}_{N}}{h^{3N}}.
\end{equation}
Therefore, the yield of $J/\psi\pi^{\pm}$ cluster in the DCPC model can be calculated by
\begin{equation}\label{eq3}
  Y_{Z_c^{\pm}(3900)\to J/\psi\pi^{\pm}}=\int\cdots\int\delta_{12}\frac{d\vec{q}_{\pi^{\pm}}d\vec{p}_{\pi^{\pm}}d\vec{q}_{J/\psi}d\vec{p}_{J/\psi}}{h^{6}}.
\end{equation}
\begin{equation}\label{eq4}
\delta_{12}=
\left\{
             \begin{array}{ll}
             1 &~ if ~1\equiv \pi^{\pm},2\equiv J/\psi; \\
               & ~m_{0}-\Delta m\leq m_{inv}\leq m_{0}+\Delta m; \\
              & ~\mid\vec{q}_{12}\mid\leq R_{0}; \\
             0 &~ \textrm{otherwise}.
             \end{array}
\right.
\end{equation}

\begin{equation}\label{eq5}
    m_{inv}=[(E_{\pi^{\pm}}+E_{J/\psi})^{2}-(p_{\pi^{\pm}}+p_{J/\psi})^{2}]^{1/2}.
\end{equation}
Where $m_{0}=3887.2$ MeV/$c^{2}$ represent the rest mass of $Z_c^{\pm}(3900)$ from PDG~\cite{r36}, and $\Delta m$ refers to its mass uncertainty. $R_0$ is the effective radius of the possible combination of $\pi^{\pm}$ and $J/\psi$ to form $Z_c^{\pm}(3900)$ and $|\vec{q}_{12}|=|\vec{q}_{1}-\vec{q}_{2}|$ represents the distance between $\pi^{\pm}$ and $J/\psi$. The $Z_c^\pm(3900)$ is constructed by the combination of hadrons $J/\psi$ and $\pi^\pm$ after the final hadrons produced with the PACIAE model. In Eq.(1), the energy function $H$ satisfies $H^2 = (\vec p_1 + \vec p_2)^2 + m_{inv}^2$ and the energy threshold satisfies $E_{a,b}^2 = (\vec p_1 + \vec p_2)^2 + (m_0 \mp \Delta m)^2$. Thus, the dynamic constraint condition $m_{0}-\Delta m\leq m_{inv}\leq m_{0}+\Delta m$ in Eq.(4) is equivalent to $E_a \leq H \leq E_b$ in Eq.(1).

\section{Calculations and results}
In the production of final states particles with PACIAE, the model parameters are fixed on the default values given in the PYTHIA model, except for the \textit{K} factor and the parameters of parj(1), parj(2), and parj(3) which are determined by fitting to the LHC data in $pp$ collisions at $\sqrt{s}=7$ TeV. Here, parj(1) is the suppression of diquark-antidiquark pair production compared with the
quark-antiquark pair production, parj(2) is the suppression of strange quark pair production compared with up (down) quark pair production, parj(3) is the extra suppression of strange diquark production compared with the normal suppression of a strange quark. We choose parj(1) $=0.10$, parj(2) $=0.20$, parj(3) $=0.90$.
To validate the production of $\pi^{\pm}$ and $J/\psi$ with PACIAE model, the yields of $\pi^{\pm}$ and $J/\psi$ are calculated with $|y|<0.5$, $0.1<p_{T}<3$ GeV/c for $\pi^{\pm}$ and $2.0<y<4.5$, $0<p_{T}<14$ GeV/c for $J/\psi$ according to LHC data separately. The results are shown in Table\ref{biao1}, together with the experimental data ~\cite {r40,r41}, which are consistent with each other within uncertainties.

\begin{table}[ht]
\centering
\caption{The yield of $\pi^{\pm}$ and $J/\psi$ in $pp$ collisions at $\sqrt{s} = 7$ TeV simulated by the PACIAE model, and compared with experimental data~\cite {r40,r41}, with the $|y|<0.5$, $0.1<p_{T}<3$ GeV/c for $\pi^{\pm}$ and $2.0<y<4.5$, $0<p_{T}<14$ GeV/c for $J/\psi$, respectively. Here, $J/\psi$ is from b decay. }
\begin{tabular}{cccc}
\hline\hline
Particle &LHC~\cite {r40,r41} & PACIAE \\ \hline
$J/\psi$ &$(1.60\pm0.01\pm0.023)\times10^{-5}$ & $(1.60\pm 0.03)\times10^{-5}$ \\
$\pi^{+}$ & $2.26\pm0.10$ & $2.26\pm 0.01$ \\
$\pi^{-}$ & $2.23\pm0.10$ & $2.25\pm 0.03$ \\ \hline
\end{tabular}\label{biao1}
\end{table}

\begin{figure*}[t]
\includegraphics[width=0.42\textwidth]{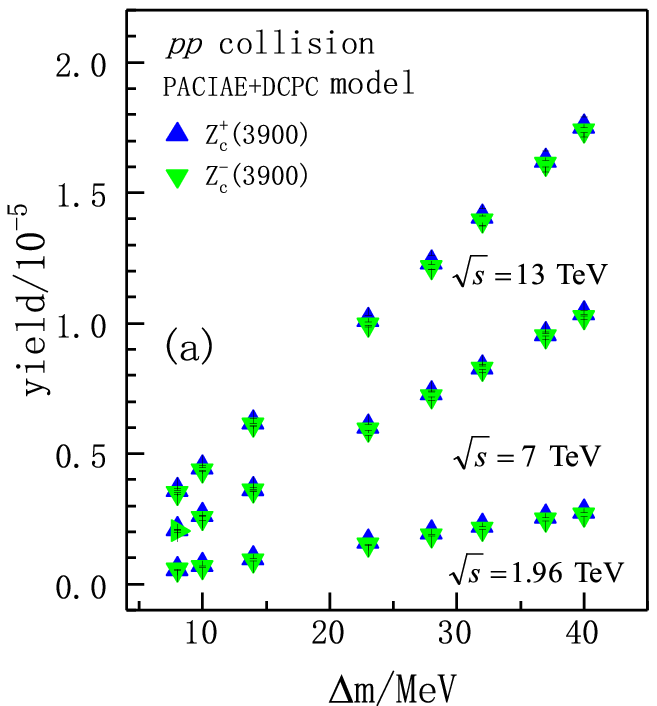}
\includegraphics[width=0.42\textwidth]{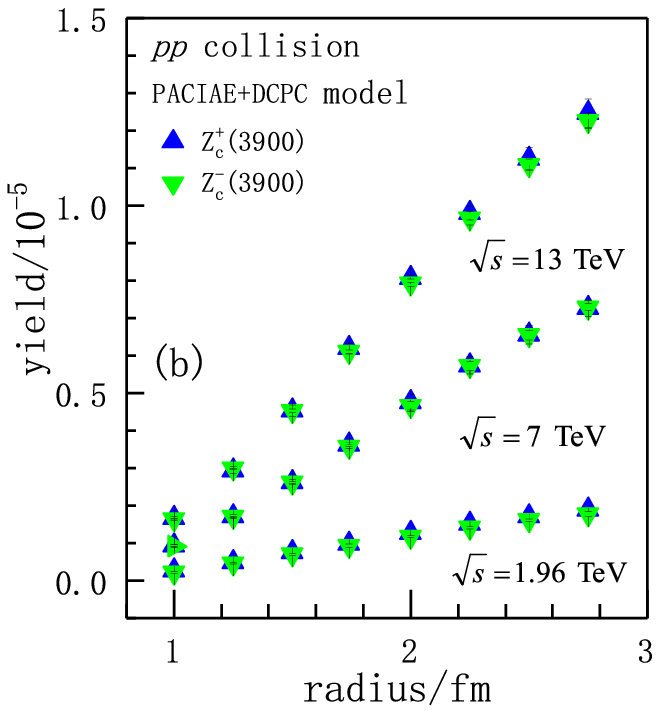}
\caption{The distribution of the yield of  exotic resonant states $Z_c^{\pm}(3900)$ in $pp$ collisions at $\sqrt{s} = 1.96, 7, 13$~TeV, respectively. (a) as a function of mass uncertainty $\Delta m$, (b) as a function of radius parameter $R_0$. The data are calculated using PACIAE+DCPC model based on the $Z_c^{\pm}(3900)\to J/\psi\pi^{\pm}$ bound state in the decay chain of $b$ hadrons. }

\end{figure*} \label{tu1}
\begin{table*}[!]
\centering
\caption{The yields $(10^{-6})$ of  exotic resonant states $Z_c^{+}(3900)$ and $Z_c^{-}(3900)$ varies with parameter $\Delta m$ changing from 8 MeV to 40 MeV in $pp$ collision at $\sqrt{s}= 1.96, 7$ and 13~TeV. $Z_c^{\pm}(3900)$ states decaying to $J/\psi\pi^{\pm}$ are computed with PACIAE + DCPC model with the radius parameter $R_0$ fixed to 1.74~fm, based on the $Z_c^{\pm}(3900)\to J/\psi\pi^{\pm}$ bound state in the decay chain of $b$ hadrons.}
\setlength{\tabcolsep}{4.5mm}
\begin{tabular}{ ccccccc }
	\hline \hline
    {$\Delta m$} & \multicolumn{2}{ c }{1.96 TeV} &  \multicolumn{2}{ c }{7 TeV}&  \multicolumn{2}{ c }{13 TeV}  \\ \cline{2-7}
	(MeV) & $Z_c^{+}(3900)$ & $Z_c^{-}(3900)$& $Z_c^{+}(3900)$& $Z_c^{-}(3900)$& $Z_c^{+}(3900)$& $Z_c^{-}(3900)$ \\ \hline
    8  & $0.57\pm0.03$ & $0.55\pm0.03$& $2.10\pm0.01$ & $2.02\pm0.05$ & $3.59\pm0.09$ & $3.51\pm0.08 $  \\
	10 & $0.72\pm0.02$ & $0.66\pm0.03$& $2.63\pm0.03$ & $2.54\pm0.07$ & $4.46\pm0.11$ & $4.38\pm0.09 $    \\
	14.1 & $0.99\pm0.07$ & $0.93\pm0.05$& $3.63\pm0.05$ & $3.58\pm0.05$ & $6.19\pm0.14$ & $6.11\pm0.05 $  \\
	23 & $1.61\pm0.10$ & $1.51\pm0.02$& $6.02\pm0.10$ & $5.92\pm0.20$ & $10.13\pm0.19$& $9.95\pm0.10 $  \\
	28 & $1.97\pm0.11$ & $1.87\pm0.04$& $7.29\pm0.10$ & $7.21\pm0.16$ & $12.32\pm0.27$& $12.16\pm0.10$ \\
	32 & $2.24\pm0.09$ & $2.14\pm0.05$& $8.30\pm0.07$ & $8.26\pm0.15$ & $14.06\pm0.32$& $13.94\pm0.18$ \\
	37 & $2.56\pm0.13$ & $2.48\pm0.06$& $9.58\pm0.08$ & $9.50\pm0.11$ & $16.21\pm0.40$& $16.13\pm0.13$ \\
	40 & $2.78\pm0.11$ & $2.66\pm0.08$& $10.38\pm0.12$& $10.25\pm0.10$& $17.52\pm0.39$& $17.40\pm0.09$ \\ \hline

\end{tabular}\label{biao2}
\end{table*}

\begin{table*}[!]
\centering
\caption{The yields $(10^{-6})$ of  exotic resonant states $Z_c^{+}(3900)$ and $Z_c^{-}(3900)$ varies with parameter radius changing from 1~fm to 2.75~fm in $pp$ collision at $\sqrt{s}= 1.96, 7$ and 13~TeV $Z_c^{\pm}(3900)$ states decaying to $J/\psi\pi^{\pm}$ are computed with PACIAE + DCPC model with the value of parameter $\Delta m$ fixed to14.1~MeV, based on the $Z_c^{\pm}(3900)\to J/\psi\pi^{\pm}$ bound state in the decay chain of $b$ hadrons.}
\setlength{\tabcolsep}{4.5mm}

\begin{tabular}{ ccccccc }
	\hline \hline
    {$R_0$} & \multicolumn{2}{ c }{1.96 TeV} &  \multicolumn{2}{ c }{7 TeV}&  \multicolumn{2}{ c }{13 TeV}  \\ \cline{2-7}
	(fm) & $Z_c^{+}(3900)$ & $Z_c^{-}(3900)$& $Z_c^{+}(3900)$& $Z_c^{-}(3900)$& $Z_c^{+}(3900)$& $Z_c^{-}(3900)$   \\ \hline
	1.00 & $0.27\pm0.04$ & $0.23\pm0.02$& $0.94\pm0.02$ & $0.92\pm0.03$ & $1.66\pm0.05$ & $1.65\pm0.02 $  \\
	1.25 & $0.48\pm0.01$ & $0.47\pm0.02$& $1.70\pm0.02$ & $1.71\pm0.04$ & $2.93\pm0.06$ & $2.99\pm0.04 $  \\
    1.50 & $0.75\pm0.02$ & $0.70\pm0.03$& $2.63\pm0.02$ & $2.62\pm0.06$ & $4.54\pm0.16$ & $4.54\pm0.05 $  \\
    1.74 & $0.99\pm0.07$ & $0.93\pm0.05$& $3.63\pm0.05$ & $3.58\pm0.05$ & $6.19\pm0.14$ & $6.11\pm0.05 $  \\
    2.00 & $1.26\pm0.02$ & $1.18\pm0.02$& $4.75\pm0.23$ & $4.67\pm0.11$ & $8.06\pm0.10$ & $7.94\pm0.10 $ \\
	2.25 & $1.50\pm0.03$ & $1.42\pm0.03$& $5.74\pm0.24$ & $5.73\pm0.12$ & $9.80\pm0.17$ & $9.67\pm0.17 $ \\
	2.50 & $1.71\pm0.01$ & $1.61\pm0.03$& $6.56\pm0.24$ & $6.55\pm0.12$ & $11.25\pm0.30$& $11.10\pm0.15$ \\
	2.75 & $1.88\pm0.01$ & $1.79\pm0.06$& $7.26\pm0.20$ & $7.29\pm0.10$ & $12.47\pm0.38$& $12.27\pm0.21$ \\ \hline
\end{tabular}\label{biao3}
\end{table*}

Then event samples with $J/\psi$, $\pi^{+}$ and $\pi^{-}$ final states are generated by PACIAE model in $pp$ collisions at $\sqrt{s} = 1.96, 7, 13$ TeV with $|y|<6$ and $0<p_T<10$~GeV/c, respectively. And the final state particles $J/\psi$ and $\pi^{\pm}$ from b-hadron decay chains are put into DCPC model to construct the $J/\psi\pi^{\pm}$ clusters, the molecular state of the $Z_c^{\pm}(3900)$.

Actually, $J/\psi$ can be originated in three different sources in $pp$ collision ~\cite {r41,r37,r38,r39}: direct prompt $J/\psi$ production, indirect prompt $J/\psi$ production, and non-prompt $J/\psi$ production from b-hadron decay chains. The sum of the first two sources is often called "prompt $J/\psi$" and the third source is called "non-prompt $J/\psi$ from b".
The ratio of prompt $J/\psi$ to non-prompt $J/\psi$ from b decay calculated by the experimental results from the LHCb in $pp$ collisions at $\sqrt{s}= 7$~TeV with $2.0 < y < 4.5$ is about 9:1. By calculating the total yield of $J/\psi$ and the yield of "non-prompt" $J/\psi$ in $pp$ at $\sqrt{s}= 7$~TeV with the b-tag method with PACIAE model under the condition of $-5.0 < y <5.0$, we can easily extract the ratio of prompt $J/\psi$ to non-prompt $J/\psi$ from b decay, which is determined to be $5.73\pm0.05$.
Here, the $Z_c^{\pm}(3900)$ are generated through the combination of $J/\psi$ and $\pi^{\pm}$ from b-hadron decay chains during the hadron evolution period.


\begin{figure}[t]
\includegraphics[width=0.4\textwidth]{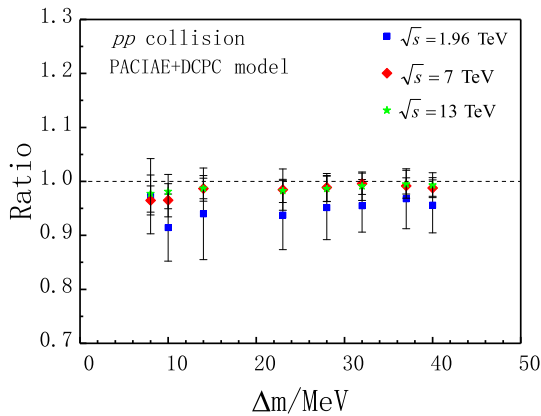}
\caption{The ratio distribution of ${Z_c^{-}(3900)}$ to ${Z_c^{+}(3900)}$ in $pp$ collisions at $\sqrt{s} = 1.96, 7$ and $13$ TeV with the value of radius parameter $R_0= 1.74$~fm, as a function of mass uncertainty $\Delta m$.}
\label{tu2}
\end{figure}
Table\ref{biao2} shows the yield of exotic state $Z_c^{+}(3900)$ and $Z_c^{-}(3900)$ in $pp$ collision at $\sqrt{s} = 1.96, 7$ and 13 TeV with parameter ${\Delta m}$ changing from 8~MeV to 40~MeV while the radius parameter is fixed to 1.74~fm. The distribution of $Z_c^{\pm}(3900)$ as a function of ${\Delta m}$ is also shown in Fig.1(a). From the Tab.\ref{biao2} and Fig.1(a), we can conclude that the yield of the exotic $Z_c^{\pm}(3900)$ states computed by PACIAE+DCPC model increases from $10^{-6}$ to $10^{-5}$in a linear way as the parameter ${\Delta m}$ increases. As the center of mass energy increases from 1.96 TeV to 13 TeV, the yield of  exotic $Z_c^{\pm}(3900)$ states calculated by PACIAE+DCPC increases.

Similarly, Table\ref{biao3} presents the yield of exotic state $Z_c^{+}(3900)$ and $Z_c^{-}(3900)$ in $pp$ collisions at $\sqrt{s} = 1.96, 7$ and 13 TeV with parameter $R$ varying from 1.0~fm to 2.75~fm at a given mass uncertainty $\Delta m=14.1$~MeV. The distribution of yield of exotic states $Z_c^{\pm}(3900)$ vs parameter $R_0$ is shown in Fig.1(b). From the Table\ref{biao3}, one can conclude that the yield of the exotic $Z_c^{\pm}(3900)$ states also increase with parameter $R_0$ from 1.0~fm to 2.75~fm at a given mass uncertainty ${\Delta m}=14.1$~MeV. But when $R_0$ is greater than 2, the distribution tends to be a little bit saturated, because the density of particle number decreases with the increase of $r$ in high energy $pp$ collisions.

\begin{figure*}[!]
\includegraphics[width=0.8\textwidth]{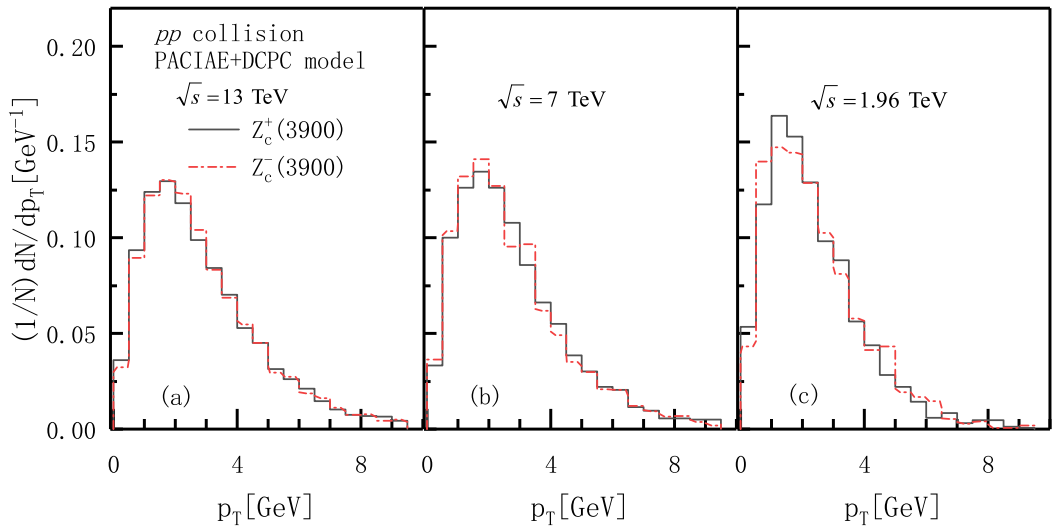}
\caption{The transverse momentum distributions of exotic state $Z_c^{+}(3900)$ (the solid line) and $Z_c^{-}(3900)$ (dashed line) calculated by {PACIAE+DCPC} model simulations with $\Delta m = 14.1$~MeV and $R_0=1.74$~fm, based on the $Z_c^{\pm}(3900)\to J/\psi\pi^{\pm}$ bound state from the decay chain of $b$ hadrons in $pp$ collision at $\sqrt{s} = 1.96, 7$ and 13 TeV, respectively.}
\label{tu3}
\end{figure*}

As a reasonable prediction, we take half of the decay width for  exotic states $Z_c^{\pm}(3900)$ in PDG~\cite {r36} as $\Delta m$ parameter, i.e, $\Delta m = \Gamma/2=14.1$~MeV, and take radius parameter $R_0=1.74$~fm~\cite{r42} relying on the analogy between the bound state ($J/\psi\pi^{\pm}$) and the structure of deuteron ($pn$). Then we may predict the yields of the  exotic states $Z_c^{\pm}(3900)$ in $pp$ collision at $\sqrt{s}=1.96, 7$, and 13~TeV, as shown in the row $R_0 = 1.74$~fm in Table\ref{biao3}. The yields of $Z_c^{\pm}(3900)$ calculated using the PACIAE+DCPC model by us are in agreement with those computed from data of Ref~\cite{r43} in the $pp/\bar p$ collisions.

So far, $Z_c^\pm(3900)$ has three possible decay modes $Z_c^\pm(3900)\to J/\psi\pi^{\pm}$, $D\bar D^*$, and $\eta_c(1s)\rho(770)^\pm$, the $\Gamma(D\bar D^*)/\Gamma(J/\psi\pi^\pm)=6.2\pm 1.1\pm 2.7$ and $\Gamma(\eta_c(1s)\rho(770)^\pm)/\Gamma(J/\psi\pi^\pm)$=2.3$\pm 0.8$ according to PDG ~\cite{r44} from BESIII experiment~\cite{r45,r46}.
Using the PACIAE model, the results we get are $\Gamma(D\bar D^*)/\Gamma(J/\psi\pi^\pm)=6.36\pm 0.02$ and $\Gamma(\eta_c(1S)\rho(770)^\pm)/\Gamma(J/\psi\pi^\pm)=1.78\pm 0.02$ respectively, which are consistent with BES\uppercase\expandafter{\romannumeral3} results. Therefore, the yield of $Z_c^\pm(3900) \to J/\psi\pi^{\pm}$ decay mode is approximately 10.9\% of the total yield of $Z_c^\pm(3900)$. So the total yield of $Z_c^\pm(3900)$ is approximately the yield of $J/\psi\pi^{\pm}$ decay times a factor of 9.1.

To facilitate the comparison between ${Z_c^{-}(3900)}$ and ${Z_c^{+}(3900)}$ in $pp$ collisions, the yield ratios of ${Z_c^{-}(3900)}$ to ${Z_c^{+}(3900)}$
computed by PACIAE+DCPC model are presented in Fig.~\ref{tu2}, which is slightly less than 1 and indicates that the production of antiparticles ${Z_c^{-}(3900)}$ is more difficult than that of particles ${Z_c^{+}(3900)}$.

\begin{figure*}[t]
\includegraphics[width=0.8\textwidth]{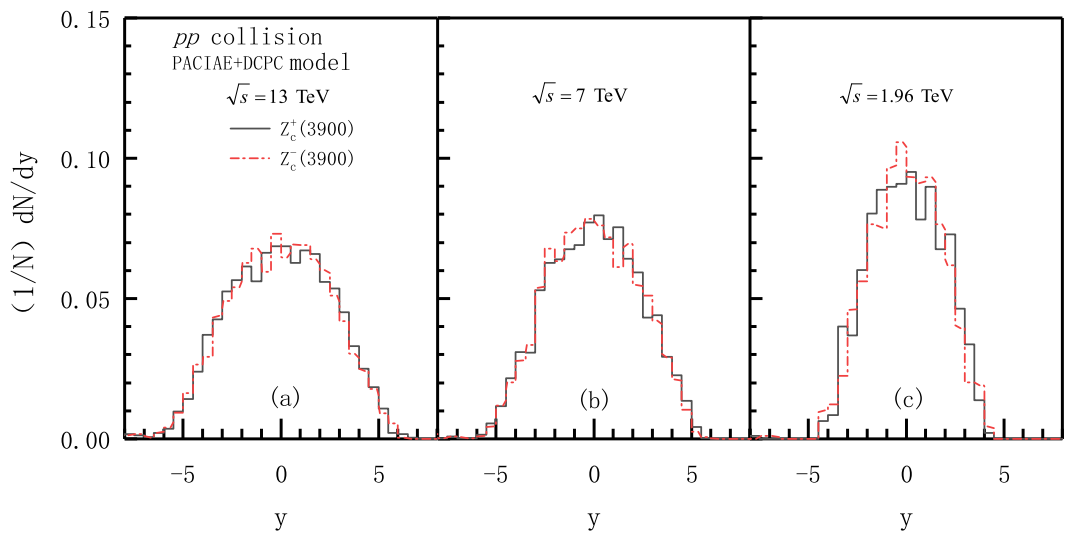}
\caption{The rapidity distributions of exotic state $Z_c^{+}(3900)$ (the solid line) and $Z_c^{-}(3900)$ (dashed line) calculated by {PACIAE+DCPC} model simulations with $\Delta m = 14.1$~MeV and $R_0=1.74$~fm, based on the $Z_c^{\pm}(3900)\to J/\psi\pi^{\pm}$ bound state from the decay chain of $b$ hadrons in $pp$ collision at $\sqrt{s} = 1.96, 7$ and 13 TeV, respectively.}\label{tu4}
\end{figure*}

The transverse momentum distribution of $Z_c^{\pm}(3900)$ calculated using PACIAE+DCPC model in $pp$ collision at $\sqrt{s}= 1.96, 7$ and 13~TeV are shown in Fig.\ref{tu3}. In each panel, the dashed line and the solid line refers to the distribution of antiparticles $Z_c^{-}(3900)$ and particles $Z_c^{+}(3900)$, respectively. Here, mass uncertainty parameter is taken as $\Delta m = \Gamma/2=14.1$~MeV~\cite {r36}, and radius parameter is taken $R_0=1.74$~fm. It can be seen from this figure that the transverse momentum distribution characteristics of antiparticles $Z_c^{-}(3900)$ is the same as that of positive particles  $Z_c^{+}(3900)$ at the same center of mass energy. But the transverse momentum distribution of the exotic resonant states $Z_c^{\pm}(3900)$ becomes wider and the peak value shifts to the right with the increase of the collision energy. The values of average transverse momentum are $2.46\pm0.16$, $2.89\pm0.06$, $3.05\pm0.04$~GeV/c for $Z_c^{+}(3900)$ and $2.50\pm0.04$, $2.80\pm0.06$, $3.06\pm0.03$~GeV/c for $Z_c^{-}(3900)$ in $pp$ collision at $\sqrt{s}= 1.96, 7$ and 13~TeV, respectively.

The rapidity distributions of $Z_c^{\pm}(3900)$ are also calculated by PACIAE+DCPC model which are shown in Fig.\ref{tu4}. It can be seen from this figure that the rapidity distribution characteristics of antiparticles $Z_c^{-}(3900)$ are the same as that of positive particles  $Z_c^{+}(3900)$ at the same center of mass energy. But the rapidity distribution of the  exotic resonant states $Z_c^{\pm}(3900)$ becomes wider with the increase of the collision energy.
\section{Summary}
\par In this paper, we study the production of $Z_c\pm(3900)$ in PACIAE+DCPC model at $\sqrt{s} = 1.96, 7$, and 13~TeV based on the $Z_c\pm(3900)\to J/\psi\pi^{\pm}$ bound state in the decay chain of $b$ hadrons. First, we study the parameter dependence of $Z_c^{\pm}(3900)$ generation on mass uncertainty $\Delta m$ from 8 to 40~MeV and radius parameters $R_0$ from 1.0 to 2.75~fm. The results indicate that the yield of $Z_c^{\pm}(3900)$ increased with the increase of parameter $\Delta m$ and $R_0$. If the parameters are chosen as $\Delta m=\Gamma/2=14.1$~MeV and $R_0 = 1.74$~fm, we can predict that the yields of $Z_c^{+}(3900)$ and $Z_c^{-}(3900)$ are $(0.99 \pm 0.07)$E-6, $(3.63 \pm 0.05)$E-6, $(6.19 \pm 0.14)$E-6, and $(0.93 \pm 0.05)$E-6, $(3.58 \pm 0.05)$E-6, $(6.11 \pm 0.05)$E-6 under three different energies of $1.96, 7, 13$~TeV in $pp$ collisions, respectively. These yields of $Z_c^{\pm}(3900)$ calculated in the PACIAE+DCPC model agree with Ref~\cite {r43}. Then, the energy dependence of rapidity and transverse momentum distribution of exotic state $Z_c^{\pm}(3900)$ are studied. The width of these distributions become larger and their peaks value get smaller with the increase of energy from 1.96~TeV to 13~TeV. In addition, it is also found that the yield ratio of antiparticle $Z_c^{-}(3900)$ to $Z_c^{+}(3900)$ is less than 1, although their distribution of rapidity and transverse momentum are the consistent in $pp$ collisions at different energies.

The study of the exotic resonant state $Z_c^{\pm}(3900)$ productions in $pp$ collisions is under way. To obtain further insight and understanding of the nature of the exotic resonant state $Z_c^{\pm}(3900)$, we therefore suggest measurements of their production rates in $pp$ and heavy-ion collisions by the LHCb experiments.

\begin{acknowledgments}
The authors thank Prof. Kang Xiao-Lin for valuable comments. This work is supported by the NSFC (11475149, 11775094, 11905188), as well as supported by the high-performance computing platform of China University of Geosciences.
\end{acknowledgments}

\bibliography{reference}

\begin{thebibliography}{44}%
\makeatletter
\providecommand \@ifxundefined [1]{%
 \@ifx{#1\undefined}
}%
\providecommand \@ifnum [1]{%
 \ifnum #1\expandafter \@firstoftwo
 \else \expandafter \@secondoftwo
 \fi
}%
\providecommand \@ifx [1]{%
 \ifx #1\expandafter \@firstoftwo
 \else \expandafter \@secondoftwo
 \fi
}%
\providecommand \natexlab [1]{#1}%
\providecommand \enquote  [1]{``#1''}%
\providecommand \bibnamefont  [1]{#1}%
\providecommand \bibfnamefont [1]{#1}%
\providecommand \citenamefont [1]{#1}%
\providecommand \href@noop [0]{\@secondoftwo}%
\providecommand \href [0]{\begingroup \@sanitize@url \@href}%
\providecommand \@href[1]{\@@startlink{#1}\@@href}%
\providecommand \@@href[1]{\endgroup#1\@@endlink}%
\providecommand \@sanitize@url [0]{\catcode `\\12\catcode `\$12\catcode
  `\&12\catcode `\#12\catcode `\^12\catcode `\_12\catcode `\%12\relax}%
\providecommand \@@startlink[1]{}%
\providecommand \@@endlink[0]{}%
\providecommand \url  [0]{\begingroup\@sanitize@url \@url }%
\providecommand \@url [1]{\endgroup\@href {#1}{\urlprefix }}%
\providecommand \urlprefix  [0]{URL }%
\providecommand \Eprint [0]{\href }%
\providecommand \doibase [0]{http://dx.doi.org/}%
\providecommand \selectlanguage [0]{\@gobble}%
\providecommand \bibinfo  [0]{\@secondoftwo}%
\providecommand \bibfield  [0]{\@secondoftwo}%
\providecommand \translation [1]{[#1]}%
\providecommand \BibitemOpen [0]{}%
\providecommand \bibitemStop [0]{}%
\providecommand \bibitemNoStop [0]{.\EOS\space}%
\providecommand \EOS [0]{\spacefactor3000\relax}%
\providecommand \BibitemShut  [1]{\csname bibitem#1\endcsname}%
\let\auto@bib@innerbib\@empty
\bibitem [{\citenamefont {Maiani}\ \emph {et~al.}(2005)\citenamefont {Maiani},
  \citenamefont {Piccinini}, \citenamefont {Polosa},\ and\ \citenamefont
  {Riquer}}]{r1}%
  \BibitemOpen
  \bibfield  {author} {\bibinfo {author} {\bibfnamefont {L.}~\bibnamefont
  {Maiani}}, \bibinfo {author} {\bibfnamefont {F.}~\bibnamefont {Piccinini}},
  \bibinfo {author} {\bibfnamefont {A.}~\bibnamefont {Polosa}}, \ and\ \bibinfo
  {author} {\bibfnamefont {V.}~\bibnamefont {Riquer}},\ }\href {\doibase
  10.1103/PhysRevD.71.014028} {\bibfield  {journal} {\bibinfo  {journal} {Phys.
  Rev. D}\ }\textbf {\bibinfo {volume} {71}},\ \bibinfo {pages} {014028}
  (\bibinfo {year} {2005})},\ \Eprint {http://arxiv.org/abs/hep-ph/0412098}
  {arXiv:hep-ph/0412098} \BibitemShut {NoStop}%
\bibitem [{\citenamefont {Maiani}\ \emph {et~al.}(2013)\citenamefont {Maiani},
  \citenamefont {Riquer}, \citenamefont {Faccini}, \citenamefont {Piccinini},
  \citenamefont {Pilloni},\ and\ \citenamefont {Polosa}}]{r2}%
  \BibitemOpen
  \bibfield  {author} {\bibinfo {author} {\bibfnamefont {L.}~\bibnamefont
  {Maiani}}, \bibinfo {author} {\bibfnamefont {V.}~\bibnamefont {Riquer}},
  \bibinfo {author} {\bibfnamefont {R.}~\bibnamefont {Faccini}}, \bibinfo
  {author} {\bibfnamefont {F.}~\bibnamefont {Piccinini}}, \bibinfo {author}
  {\bibfnamefont {A.}~\bibnamefont {Pilloni}}, \ and\ \bibinfo {author}
  {\bibfnamefont {A.}~\bibnamefont {Polosa}},\ }\href {\doibase
  10.1103/PhysRevD.87.111102} {\bibfield  {journal} {\bibinfo  {journal} {Phys.
  Rev. D}\ }\textbf {\bibinfo {volume} {87}},\ \bibinfo {pages} {111102}
  (\bibinfo {year} {2013})},\ \Eprint {http://arxiv.org/abs/1303.6857}
  {arXiv:1303.6857 [hep-ph]} \BibitemShut {NoStop}%
\bibitem [{\citenamefont {Karliner}\ and\ \citenamefont
  {Nussinov}(2013)}]{r12}%
  \BibitemOpen
  \bibfield  {author} {\bibinfo {author} {\bibfnamefont {M.}~\bibnamefont
  {Karliner}}\ and\ \bibinfo {author} {\bibfnamefont {S.}~\bibnamefont
  {Nussinov}},\ }\href {\doibase 10.1007/JHEP07(2013)153} {\bibfield  {journal}
  {\bibinfo  {journal} {JHEP}\ }\textbf {\bibinfo {volume} {07}},\ \bibinfo
  {pages} {153} (\bibinfo {year} {2013})},\ \Eprint
  {http://arxiv.org/abs/1304.0345} {arXiv:1304.0345 [hep-ph]} \BibitemShut
  {NoStop}%
\bibitem [{\citenamefont {Guo}\ \emph {et~al.}(2013)\citenamefont {Guo},
  \citenamefont {Hidalgo-Duque}, \citenamefont {Nieves},\ and\ \citenamefont
  {Valderrama}}]{r3}%
  \BibitemOpen
  \bibfield  {author} {\bibinfo {author} {\bibfnamefont {F.-K.}\ \bibnamefont
  {Guo}}, \bibinfo {author} {\bibfnamefont {C.}~\bibnamefont {Hidalgo-Duque}},
  \bibinfo {author} {\bibfnamefont {J.}~\bibnamefont {Nieves}}, \ and\ \bibinfo
  {author} {\bibfnamefont {M.~P.}\ \bibnamefont {Valderrama}},\ }\href
  {\doibase 10.1103/PhysRevD.88.054007} {\bibfield  {journal} {\bibinfo
  {journal} {Phys. Rev. D}\ }\textbf {\bibinfo {volume} {88}},\ \bibinfo
  {pages} {054007} (\bibinfo {year} {2013})},\ \Eprint
  {http://arxiv.org/abs/1303.6608} {arXiv:1303.6608 [hep-ph]} \BibitemShut
  {NoStop}%
\bibitem [{\citenamefont {Cui}\ \emph {et~al.}(2014)\citenamefont {Cui},
  \citenamefont {Liu}, \citenamefont {Chen},\ and\ \citenamefont {Huang}}]{r4}%
  \BibitemOpen
  \bibfield  {author} {\bibinfo {author} {\bibfnamefont {C.-Y.}\ \bibnamefont
  {Cui}}, \bibinfo {author} {\bibfnamefont {Y.-L.}\ \bibnamefont {Liu}},
  \bibinfo {author} {\bibfnamefont {W.-B.}\ \bibnamefont {Chen}}, \ and\
  \bibinfo {author} {\bibfnamefont {M.-Q.}\ \bibnamefont {Huang}},\ }\href
  {\doibase 10.1088/0954-3899/41/7/075003} {\bibfield  {journal} {\bibinfo
  {journal} {J. Phys. G}\ }\textbf {\bibinfo {volume} {41}},\ \bibinfo {pages}
  {075003} (\bibinfo {year} {2014})},\ \Eprint {http://arxiv.org/abs/1304.1850}
  {arXiv:1304.1850 [hep-ph]} \BibitemShut {NoStop}%
\bibitem [{\citenamefont {Zhang}(2013)}]{r5}%
  \BibitemOpen
  \bibfield  {author} {\bibinfo {author} {\bibfnamefont {J.-R.}\ \bibnamefont
  {Zhang}},\ }\href {\doibase 10.1103/PhysRevD.87.116004} {\bibfield  {journal}
  {\bibinfo  {journal} {Phys. Rev. D}\ }\textbf {\bibinfo {volume} {87}},\
  \bibinfo {pages} {116004} (\bibinfo {year} {2013})},\ \Eprint
  {http://arxiv.org/abs/1304.5748} {arXiv:1304.5748 [hep-ph]} \BibitemShut
  {NoStop}%
\bibitem [{\citenamefont {Braaten}(2013)}]{r6}%
  \BibitemOpen
  \bibfield  {author} {\bibinfo {author} {\bibfnamefont {E.}~\bibnamefont
  {Braaten}},\ }\href {\doibase 10.1103/PhysRevLett.111.162003} {\bibfield
  {journal} {\bibinfo  {journal} {Phys. Rev. Lett.}\ }\textbf {\bibinfo
  {volume} {111}},\ \bibinfo {pages} {162003} (\bibinfo {year} {2013})},\
  \Eprint {http://arxiv.org/abs/1305.6905} {arXiv:1305.6905 [hep-ph]}
  \BibitemShut {NoStop}%
\bibitem [{\citenamefont {Close}\ and\ \citenamefont {Page}(2005)}]{r16}%
  \BibitemOpen
  \bibfield  {author} {\bibinfo {author} {\bibfnamefont {F.~E.}\ \bibnamefont
  {Close}}\ and\ \bibinfo {author} {\bibfnamefont {P.~R.}\ \bibnamefont
  {Page}},\ }\href {\doibase 10.1016/j.physletb.2005.09.016} {\bibfield
  {journal} {\bibinfo  {journal} {Phys. Lett. B}\ }\textbf {\bibinfo {volume}
  {628}},\ \bibinfo {pages} {215} (\bibinfo {year} {2005})},\ \Eprint
  {http://arxiv.org/abs/hep-ph/0507199} {arXiv:hep-ph/0507199} \BibitemShut
  {NoStop}%
\bibitem [{\citenamefont {Ablikim}\ \emph {et~al.}(2013)\citenamefont {Ablikim}
  \emph {et~al.}}]{r7}%
  \BibitemOpen
  \bibfield  {author} {\bibinfo {author} {\bibfnamefont {M.}~\bibnamefont
  {Ablikim}} \emph {et~al.} (\bibinfo {collaboration} {BESIII}),\ }\href
  {\doibase 10.1103/PhysRevLett.110.252001} {\bibfield  {journal} {\bibinfo
  {journal} {Phys. Rev. Lett.}\ }\textbf {\bibinfo {volume} {110}},\ \bibinfo
  {pages} {252001} (\bibinfo {year} {2013})},\ \Eprint
  {http://arxiv.org/abs/1303.5949} {arXiv:1303.5949 [hep-ex]} \BibitemShut
  {NoStop}%
\bibitem [{\citenamefont {Ablikim}\ \emph {et~al.}(2010)\citenamefont {Ablikim}
  \emph {et~al.}}]{r11}%
  \BibitemOpen
  \bibfield  {author} {\bibinfo {author} {\bibfnamefont {M.}~\bibnamefont
  {Ablikim}} \emph {et~al.} (\bibinfo {collaboration} {BESIII}),\ }\href
  {\doibase 10.1016/j.nima.2009.12.050} {\bibfield  {journal} {\bibinfo
  {journal} {Nucl. Instrum. Meth. A}\ }\textbf {\bibinfo {volume} {614}},\
  \bibinfo {pages} {345} (\bibinfo {year} {2010})},\ \Eprint
  {http://arxiv.org/abs/0911.4960} {arXiv:0911.4960 [physics.ins-det]}
  \BibitemShut {NoStop}%
\bibitem [{\citenamefont {Ablikim}\ \emph {et~al.}(2015)\citenamefont {Ablikim}
  \emph {et~al.}}]{r13}%
  \BibitemOpen
  \bibfield  {author} {\bibinfo {author} {\bibfnamefont {M.}~\bibnamefont
  {Ablikim}} \emph {et~al.} (\bibinfo {collaboration} {BESIII}),\ }\href
  {\doibase 10.1103/PhysRevLett.115.112003} {\bibfield  {journal} {\bibinfo
  {journal} {Phys. Rev. Lett.}\ }\textbf {\bibinfo {volume} {115}},\ \bibinfo
  {pages} {112003} (\bibinfo {year} {2015})},\ \Eprint
  {http://arxiv.org/abs/1506.06018} {arXiv:1506.06018 [hep-ex]} \BibitemShut
  {NoStop}%
\bibitem [{\citenamefont {Wang}(2018)}]{r14}%
  \BibitemOpen
  \bibfield  {author} {\bibinfo {author} {\bibfnamefont {Z.-G.}\ \bibnamefont
  {Wang}},\ }\href {\doibase 10.1140/epjc/s10052-018-5794-0} {\bibfield
  {journal} {\bibinfo  {journal} {Eur. Phys. J. C}\ }\textbf {\bibinfo {volume}
  {78}},\ \bibinfo {pages} {297} (\bibinfo {year} {2018})},\ \Eprint
  {http://arxiv.org/abs/1712.05664} {arXiv:1712.05664 [hep-ph]} \BibitemShut
  {NoStop}%
\bibitem [{\citenamefont {Ablikim}\ \emph
  {et~al.}(2014{\natexlab{a}})\citenamefont {Ablikim} \emph {et~al.}}]{r15}%
  \BibitemOpen
  \bibfield  {author} {\bibinfo {author} {\bibfnamefont {M.}~\bibnamefont
  {Ablikim}} \emph {et~al.} (\bibinfo {collaboration} {BESIII}),\ }\href
  {\doibase 10.1103/PhysRevLett.112.022001} {\bibfield  {journal} {\bibinfo
  {journal} {Phys. Rev. Lett.}\ }\textbf {\bibinfo {volume} {112}},\ \bibinfo
  {pages} {022001} (\bibinfo {year} {2014}{\natexlab{a}})},\ \Eprint
  {http://arxiv.org/abs/1310.1163} {arXiv:1310.1163 [hep-ex]} \BibitemShut
  {NoStop}%
\bibitem [{\citenamefont {Xiao}\ \emph {et~al.}(2013)\citenamefont {Xiao},
  \citenamefont {Dobbs}, \citenamefont {Tomaradze},\ and\ \citenamefont
  {Seth}}]{r17}%
  \BibitemOpen
  \bibfield  {author} {\bibinfo {author} {\bibfnamefont {T.}~\bibnamefont
  {Xiao}}, \bibinfo {author} {\bibfnamefont {S.}~\bibnamefont {Dobbs}},
  \bibinfo {author} {\bibfnamefont {A.}~\bibnamefont {Tomaradze}}, \ and\
  \bibinfo {author} {\bibfnamefont {K.~K.}\ \bibnamefont {Seth}},\ }\href
  {\doibase 10.1016/j.physletb.2013.10.041} {\bibfield  {journal} {\bibinfo
  {journal} {Phys. Lett. B}\ }\textbf {\bibinfo {volume} {727}},\ \bibinfo
  {pages} {366} (\bibinfo {year} {2013})},\ \Eprint
  {http://arxiv.org/abs/1304.3036} {arXiv:1304.3036 [hep-ex]} \BibitemShut
  {NoStop}%
\bibitem [{\citenamefont {Ablikim}\ \emph {et~al.}(2017)\citenamefont {Ablikim}
  \emph {et~al.}}]{r8}%
  \BibitemOpen
  \bibfield  {author} {\bibinfo {author} {\bibfnamefont {M.}~\bibnamefont
  {Ablikim}} \emph {et~al.} (\bibinfo {collaboration} {BESIII}),\ }\href
  {\doibase 10.1103/PhysRevLett.119.072001} {\bibfield  {journal} {\bibinfo
  {journal} {Phys. Rev. Lett.}\ }\textbf {\bibinfo {volume} {119}},\ \bibinfo
  {pages} {072001} (\bibinfo {year} {2017})},\ \Eprint
  {http://arxiv.org/abs/1706.04100} {arXiv:1706.04100 [hep-ex]} \BibitemShut
  {NoStop}%
\bibitem [{\citenamefont {Liu}\ \emph {et~al.}(2013)\citenamefont {Liu} \emph
  {et~al.}}]{r9}%
  \BibitemOpen
  \bibfield  {author} {\bibinfo {author} {\bibfnamefont {Z.}~\bibnamefont
  {Liu}} \emph {et~al.} (\bibinfo {collaboration} {Belle}),\ }\href {\doibase
  10.1103/PhysRevLett.110.252002} {\bibfield  {journal} {\bibinfo  {journal}
  {Phys. Rev. Lett.}\ }\textbf {\bibinfo {volume} {110}},\ \bibinfo {pages}
  {252002} (\bibinfo {year} {2013})},\ \bibinfo {note} {[Erratum:
  Phys.Rev.Lett. 111, 019901 (2013)]},\ \Eprint
  {http://arxiv.org/abs/1304.0121} {arXiv:1304.0121 [hep-ex]} \BibitemShut
  {NoStop}%
\bibitem [{\citenamefont {Casey}\ \emph {et~al.}(2013)\citenamefont {Casey}
  \emph {et~al.}}]{r18}%
  \BibitemOpen
  \bibfield  {author} {\bibinfo {author} {\bibfnamefont {B.}~\bibnamefont
  {Casey}} \emph {et~al.},\ }\href {\doibase 10.1016/j.nima.2012.08.095}
  {\bibfield  {journal} {\bibinfo  {journal} {Nucl. Instrum. Meth. A}\ }\textbf
  {\bibinfo {volume} {698}},\ \bibinfo {pages} {208} (\bibinfo {year}
  {2013})},\ \Eprint {http://arxiv.org/abs/1204.0461} {arXiv:1204.0461
  [hep-ex]} \BibitemShut {NoStop}%
\bibitem [{\citenamefont {Tuchming}(2017)}]{r23}%
  \BibitemOpen
  \bibfield  {author} {\bibinfo {author} {\bibfnamefont {B.}~\bibnamefont
  {Tuchming}},\ }\href {\doibase 10.1393/ncc/i2017-17178-2} {\bibfield
  {journal} {\bibinfo  {journal} {Nuovo Cim. C}\ }\textbf {\bibinfo {volume}
  {40}},\ \bibinfo {pages} {178} (\bibinfo {year} {2017})},\ \Eprint
  {http://arxiv.org/abs/1705.07000} {arXiv:1705.07000 [hep-ex]} \BibitemShut
  {NoStop}%
\bibitem [{\citenamefont {Abazov}\ \emph {et~al.}(2006)\citenamefont {Abazov}
  \emph {et~al.}}]{r24}%
  \BibitemOpen
  \bibfield  {author} {\bibinfo {author} {\bibfnamefont {V.}~\bibnamefont
  {Abazov}} \emph {et~al.} (\bibinfo {collaboration} {D0}),\ }\href {\doibase
  10.1016/j.nima.2006.05.248} {\bibfield  {journal} {\bibinfo  {journal} {Nucl.
  Instrum. Meth. A}\ }\textbf {\bibinfo {volume} {565}},\ \bibinfo {pages}
  {463} (\bibinfo {year} {2006})},\ \Eprint
  {http://arxiv.org/abs/physics/0507191} {arXiv:physics/0507191} \BibitemShut
  {NoStop}%
\bibitem [{\citenamefont {Abazov}\ \emph {et~al.}(2018)\citenamefont {Abazov}
  \emph {et~al.}}]{r19}%
  \BibitemOpen
  \bibfield  {author} {\bibinfo {author} {\bibfnamefont {V.~M.}\ \bibnamefont
  {Abazov}} \emph {et~al.} (\bibinfo {collaboration} {D0}),\ }\href {\doibase
  10.1103/PhysRevD.98.052010} {\bibfield  {journal} {\bibinfo  {journal} {Phys.
  Rev. D}\ }\textbf {\bibinfo {volume} {98}},\ \bibinfo {pages} {052010}
  (\bibinfo {year} {2018})},\ \Eprint {http://arxiv.org/abs/1807.00183}
  {arXiv:1807.00183 [hep-ex]} \BibitemShut {NoStop}%
\bibitem [{\citenamefont {Abazov}\ \emph {et~al.}(2019)\citenamefont {Abazov}
  \emph {et~al.}}]{r20}%
  \BibitemOpen
  \bibfield  {author} {\bibinfo {author} {\bibfnamefont {V.~M.}\ \bibnamefont
  {Abazov}} \emph {et~al.} (\bibinfo {collaboration} {D0}),\ }\href {\doibase
  10.1103/PhysRevD.100.012005} {\bibfield  {journal} {\bibinfo  {journal}
  {Phys. Rev. D}\ }\textbf {\bibinfo {volume} {100}},\ \bibinfo {pages}
  {012005} (\bibinfo {year} {2019})},\ \Eprint
  {http://arxiv.org/abs/1905.13704} {arXiv:1905.13704 [hep-ex]} \BibitemShut
  {NoStop}%
\bibitem [{\citenamefont {Aubert}\ \emph {et~al.}(1974)\citenamefont {Aubert}
  \emph {et~al.}}]{r21}%
  \BibitemOpen
  \bibfield  {author} {\bibinfo {author} {\bibfnamefont {J.}~\bibnamefont
  {Aubert}} \emph {et~al.} (\bibinfo {collaboration} {E598}),\ }\href {\doibase
  10.1103/PhysRevLett.33.1404} {\bibfield  {journal} {\bibinfo  {journal}
  {Phys. Rev. Lett.}\ }\textbf {\bibinfo {volume} {33}},\ \bibinfo {pages}
  {1404} (\bibinfo {year} {1974})}\BibitemShut {NoStop}%
\bibitem [{\citenamefont {Augustin}\ \emph {et~al.}(1974)\citenamefont
  {Augustin} \emph {et~al.}}]{r22}%
  \BibitemOpen
  \bibfield  {author} {\bibinfo {author} {\bibfnamefont {J.}~\bibnamefont
  {Augustin}} \emph {et~al.} (\bibinfo {collaboration} {SLAC-SP-017}),\ }\href
  {\doibase 10.1103/PhysRevLett.33.1406} {\bibfield  {journal} {\bibinfo
  {journal} {Phys. Rev. Lett.}\ }\textbf {\bibinfo {volume} {33}},\ \bibinfo
  {pages} {1406} (\bibinfo {year} {1974})}\BibitemShut {NoStop}%
\bibitem [{\citenamefont {Sa}\ \emph {et~al.}(2012)\citenamefont {Sa},
  \citenamefont {Zhou}, \citenamefont {Yan}, \citenamefont {Li}, \citenamefont
  {Feng}, \citenamefont {Dong},\ and\ \citenamefont {Cai}}]{r25}%
  \BibitemOpen
  \bibfield  {author} {\bibinfo {author} {\bibfnamefont {B.-H.}\ \bibnamefont
  {Sa}}, \bibinfo {author} {\bibfnamefont {D.-M.}\ \bibnamefont {Zhou}},
  \bibinfo {author} {\bibfnamefont {Y.-L.}\ \bibnamefont {Yan}}, \bibinfo
  {author} {\bibfnamefont {X.-M.}\ \bibnamefont {Li}}, \bibinfo {author}
  {\bibfnamefont {S.-Q.}\ \bibnamefont {Feng}}, \bibinfo {author}
  {\bibfnamefont {B.-G.}\ \bibnamefont {Dong}}, \ and\ \bibinfo {author}
  {\bibfnamefont {X.}~\bibnamefont {Cai}},\ }\href {\doibase
  10.1016/j.cpc.2011.08.021} {\bibfield  {journal} {\bibinfo  {journal}
  {Comput. Phys. Commun.}\ }\textbf {\bibinfo {volume} {183}},\ \bibinfo
  {pages} {333} (\bibinfo {year} {2012})},\ \Eprint
  {http://arxiv.org/abs/1104.1238} {arXiv:1104.1238 [nucl-th]} \BibitemShut
  {NoStop}%
\bibitem [{\citenamefont {Chen}\ \emph {et~al.}(2012)\citenamefont {Chen},
  \citenamefont {Yan}, \citenamefont {Li}, \citenamefont {Zhou}, \citenamefont
  {Wang}, \citenamefont {Dong},\ and\ \citenamefont {Sa}}]{r31}%
  \BibitemOpen
  \bibfield  {author} {\bibinfo {author} {\bibfnamefont {G.}~\bibnamefont
  {Chen}}, \bibinfo {author} {\bibfnamefont {Y.-L.}\ \bibnamefont {Yan}},
  \bibinfo {author} {\bibfnamefont {D.-S.}\ \bibnamefont {Li}}, \bibinfo
  {author} {\bibfnamefont {D.-M.}\ \bibnamefont {Zhou}}, \bibinfo {author}
  {\bibfnamefont {M.-J.}\ \bibnamefont {Wang}}, \bibinfo {author}
  {\bibfnamefont {B.-G.}\ \bibnamefont {Dong}}, \ and\ \bibinfo {author}
  {\bibfnamefont {B.-H.}\ \bibnamefont {Sa}},\ }\href {\doibase
  10.1103/PhysRevC.86.054910} {\bibfield  {journal} {\bibinfo  {journal} {Phys.
  Rev. C}\ }\textbf {\bibinfo {volume} {86}},\ \bibinfo {pages} {054910}
  (\bibinfo {year} {2012})},\ \Eprint {http://arxiv.org/abs/1209.4182}
  {arXiv:1209.4182 [nucl-th]} \BibitemShut {NoStop}%
\bibitem [{\citenamefont {Sjostrand}\ \emph {et~al.}(2006)\citenamefont
  {Sjostrand}, \citenamefont {Mrenna},\ and\ \citenamefont {Skands}}]{r26}%
  \BibitemOpen
  \bibfield  {author} {\bibinfo {author} {\bibfnamefont {T.}~\bibnamefont
  {Sjostrand}}, \bibinfo {author} {\bibfnamefont {S.}~\bibnamefont {Mrenna}}, \
  and\ \bibinfo {author} {\bibfnamefont {P.~Z.}\ \bibnamefont {Skands}},\
  }\href {\doibase 10.1088/1126-6708/2006/05/026} {\bibfield  {journal}
  {\bibinfo  {journal} {JHEP}\ }\textbf {\bibinfo {volume} {05}},\ \bibinfo
  {pages} {026} (\bibinfo {year} {2006})},\ \Eprint
  {http://arxiv.org/abs/hep-ph/0603175} {arXiv:hep-ph/0603175} \BibitemShut
  {NoStop}%
\bibitem [{\citenamefont {Sa}\ and\ \citenamefont {Tai}(1995)}]{r28}%
  \BibitemOpen
  \bibfield  {author} {\bibinfo {author} {\bibfnamefont {B.}~\bibnamefont
  {Sa}}\ and\ \bibinfo {author} {\bibfnamefont {A.}~\bibnamefont {Tai}},\
  }\href {\doibase 10.1016/0010-4655(95)00066-O} {\bibfield  {journal}
  {\bibinfo  {journal} {Comput. Phys. Commun.}\ }\textbf {\bibinfo {volume}
  {90}},\ \bibinfo {pages} {121} (\bibinfo {year} {1995})}\BibitemShut
  {NoStop}%
\bibitem [{\citenamefont {Zhilei}\ \emph {et~al.}(2016)\citenamefont {Zhilei},
  \citenamefont {Gang}, \citenamefont {Hongge}, \citenamefont {Tingting},\ and\
  \citenamefont {Dikai}}]{r29}%
  \BibitemOpen
  \bibfield  {author} {\bibinfo {author} {\bibfnamefont {S.}~\bibnamefont
  {Zhilei}}, \bibinfo {author} {\bibfnamefont {C.}~\bibnamefont {Gang}},
  \bibinfo {author} {\bibfnamefont {X.}~\bibnamefont {Hongge}}, \bibinfo
  {author} {\bibfnamefont {Z.}~\bibnamefont {Tingting}}, \ and\ \bibinfo
  {author} {\bibfnamefont {L.}~\bibnamefont {Dikai}},\ }\href {\doibase
  10.1140/epja/i2016-16093-2} {\bibfield  {journal} {\bibinfo  {journal} {Eur.
  Phys. J. A}\ }\textbf {\bibinfo {volume} {52}},\ \bibinfo {pages} {93}
  (\bibinfo {year} {2016})}\BibitemShut {NoStop}%
\bibitem [{\citenamefont {Chen}\ \emph {et~al.}(2014)\citenamefont {Chen},
  \citenamefont {Chen}, \citenamefont {Wang},\ and\ \citenamefont
  {Chen}}]{r30}%
  \BibitemOpen
  \bibfield  {author} {\bibinfo {author} {\bibfnamefont {G.}~\bibnamefont
  {Chen}}, \bibinfo {author} {\bibfnamefont {H.}~\bibnamefont {Chen}}, \bibinfo
  {author} {\bibfnamefont {J.-L.}\ \bibnamefont {Wang}}, \ and\ \bibinfo
  {author} {\bibfnamefont {Z.-Y.}\ \bibnamefont {Chen}},\ }\href {\doibase
  10.1088/0954-3899/41/11/115102} {\bibfield  {journal} {\bibinfo  {journal}
  {J. Phys. G}\ }\textbf {\bibinfo {volume} {41}},\ \bibinfo {pages} {115102}
  (\bibinfo {year} {2014})},\ \Eprint {http://arxiv.org/abs/1401.6872}
  {arXiv:1401.6872 [hep-ph]} \BibitemShut {NoStop}%
\bibitem [{\citenamefont {Chen}\ \emph {et~al.}(2013)\citenamefont {Chen},
  \citenamefont {Chen}, \citenamefont {Wu}, \citenamefont {Li},\ and\
  \citenamefont {Wang}}]{r32}%
  \BibitemOpen
  \bibfield  {author} {\bibinfo {author} {\bibfnamefont {G.}~\bibnamefont
  {Chen}}, \bibinfo {author} {\bibfnamefont {H.}~\bibnamefont {Chen}}, \bibinfo
  {author} {\bibfnamefont {J.}~\bibnamefont {Wu}}, \bibinfo {author}
  {\bibfnamefont {D.-S.}\ \bibnamefont {Li}}, \ and\ \bibinfo {author}
  {\bibfnamefont {M.-J.}\ \bibnamefont {Wang}},\ }\href {\doibase
  10.1103/PhysRevC.88.034908} {\bibfield  {journal} {\bibinfo  {journal} {Phys.
  Rev. C}\ }\textbf {\bibinfo {volume} {88}},\ \bibinfo {pages} {034908}
  (\bibinfo {year} {2013})},\ \Eprint {http://arxiv.org/abs/1307.4515}
  {arXiv:1307.4515 [nucl-th]} \BibitemShut {NoStop}%
\bibitem [{\citenamefont {Dong}\ \emph {et~al.}(2018)\citenamefont {Dong},
  \citenamefont {Chen}, \citenamefont {Wang}, \citenamefont {She},
  \citenamefont {Yan}, \citenamefont {Liu}, \citenamefont {Zhou},\ and\
  \citenamefont {Sa}}]{r33}%
  \BibitemOpen
  \bibfield  {author} {\bibinfo {author} {\bibfnamefont {Z.-J.}\ \bibnamefont
  {Dong}}, \bibinfo {author} {\bibfnamefont {G.}~\bibnamefont {Chen}}, \bibinfo
  {author} {\bibfnamefont {Q.-Y.}\ \bibnamefont {Wang}}, \bibinfo {author}
  {\bibfnamefont {Z.-L.}\ \bibnamefont {She}}, \bibinfo {author} {\bibfnamefont
  {Y.-L.}\ \bibnamefont {Yan}}, \bibinfo {author} {\bibfnamefont {F.-X.}\
  \bibnamefont {Liu}}, \bibinfo {author} {\bibfnamefont {D.-M.}\ \bibnamefont
  {Zhou}}, \ and\ \bibinfo {author} {\bibfnamefont {B.-H.}\ \bibnamefont
  {Sa}},\ }\href {\doibase 10.1140/epja/i2018-12580-8} {\bibfield  {journal}
  {\bibinfo  {journal} {Eur. Phys. J. A}\ }\textbf {\bibinfo {volume} {54}},\
  \bibinfo {pages} {144} (\bibinfo {year} {2018})},\ \Eprint
  {http://arxiv.org/abs/1803.01547} {arXiv:1803.01547 [nucl-th]} \BibitemShut
  {NoStop}%
\bibitem [{\citenamefont {Sittiketkorn}\ \emph {et~al.}(2017)\citenamefont
  {Sittiketkorn}, \citenamefont {Tomuang}, \citenamefont {Srisawad},
  \citenamefont {Limphirat}, \citenamefont {Herold}, \citenamefont {Yan},
  \citenamefont {Chen}, \citenamefont {Zhou}, \citenamefont {Kobdaj},\ and\
  \citenamefont {Yan}}]{r34}%
  \BibitemOpen
  \bibfield  {author} {\bibinfo {author} {\bibfnamefont {P.}~\bibnamefont
  {Sittiketkorn}}, \bibinfo {author} {\bibfnamefont {K.}~\bibnamefont
  {Tomuang}}, \bibinfo {author} {\bibfnamefont {P.}~\bibnamefont {Srisawad}},
  \bibinfo {author} {\bibfnamefont {A.}~\bibnamefont {Limphirat}}, \bibinfo
  {author} {\bibfnamefont {C.}~\bibnamefont {Herold}}, \bibinfo {author}
  {\bibfnamefont {Y.-L.}\ \bibnamefont {Yan}}, \bibinfo {author} {\bibfnamefont
  {G.}~\bibnamefont {Chen}}, \bibinfo {author} {\bibfnamefont {D.-M.}\
  \bibnamefont {Zhou}}, \bibinfo {author} {\bibfnamefont {C.}~\bibnamefont
  {Kobdaj}}, \ and\ \bibinfo {author} {\bibfnamefont {Y.}~\bibnamefont {Yan}},\
  }\href {\doibase 10.1103/PhysRevC.96.064002} {\bibfield  {journal} {\bibinfo
  {journal} {Phys. Rev. C}\ }\textbf {\bibinfo {volume} {96}},\ \bibinfo
  {pages} {064002} (\bibinfo {year} {2017})}\BibitemShut {NoStop}%
\bibitem [{\citenamefont {Stowe}(2007)}]{r35}%
  \BibitemOpen
  \bibfield  {author} {\bibinfo {author} {\bibfnamefont {K.}~\bibnamefont
  {Stowe}},\ }\href@noop {} {\emph {\bibinfo {title} {Introduction to
  Thermodynamics and Statistical Mechanics}}}\ (\bibinfo  {publisher}
  {Cambridge University, Cambridge, England},\ \bibinfo {year}
  {2007})\BibitemShut {NoStop}%
\bibitem [{\citenamefont {Tanabashi}\ \emph {et~al.}(2018)\citenamefont
  {Tanabashi} \emph {et~al.}}]{r36}%
  \BibitemOpen
  \bibfield  {author} {\bibinfo {author} {\bibfnamefont {M.}~\bibnamefont
  {Tanabashi}} \emph {et~al.} (\bibinfo {collaboration} {Particle Data
  Group}),\ }\href {\doibase 10.1103/PhysRevD.98.030001} {\bibfield  {journal}
  {\bibinfo  {journal} {Phys. Rev. D}\ }\textbf {\bibinfo {volume} {98}},\
  \bibinfo {pages} {030001} (\bibinfo {year} {2018})}\BibitemShut {NoStop}%
\bibitem [{\citenamefont {Adam}\ \emph {et~al.}(2015)\citenamefont {Adam} \emph
  {et~al.}}]{r40}%
  \BibitemOpen
  \bibfield  {author} {\bibinfo {author} {\bibfnamefont {J.}~\bibnamefont
  {Adam}} \emph {et~al.} (\bibinfo {collaboration} {ALICE}),\ }\href {\doibase
  10.1140/epjc/s10052-015-3422-9} {\bibfield  {journal} {\bibinfo  {journal}
  {Eur. Phys. J. C}\ }\textbf {\bibinfo {volume} {75}},\ \bibinfo {pages} {226}
  (\bibinfo {year} {2015})},\ \Eprint {http://arxiv.org/abs/1504.00024}
  {arXiv:1504.00024 [nucl-ex]} \BibitemShut {NoStop}%
\bibitem [{\citenamefont {Aaij}\ \emph {et~al.}(2011)\citenamefont {Aaij} \emph
  {et~al.}}]{r41}%
  \BibitemOpen
  \bibfield  {author} {\bibinfo {author} {\bibfnamefont {R.}~\bibnamefont
  {Aaij}} \emph {et~al.} (\bibinfo {collaboration} {LHCb}),\ }\href {\doibase
  10.1140/epjc/s10052-011-1645-y} {\bibfield  {journal} {\bibinfo  {journal}
  {Eur. Phys. J. C}\ }\textbf {\bibinfo {volume} {71}},\ \bibinfo {pages}
  {1645} (\bibinfo {year} {2011})},\ \Eprint {http://arxiv.org/abs/1103.0423}
  {arXiv:1103.0423 [hep-ex]} \BibitemShut {NoStop}%
\bibitem [{\citenamefont {Albajar}\ \emph {et~al.}(1991)\citenamefont {Albajar}
  \emph {et~al.}}]{r37}%
  \BibitemOpen
  \bibfield  {author} {\bibinfo {author} {\bibfnamefont {C.}~\bibnamefont
  {Albajar}} \emph {et~al.} (\bibinfo {collaboration} {UA1}),\ }\href {\doibase
  10.1016/0370-2693(91)90228-I} {\bibfield  {journal} {\bibinfo  {journal}
  {Phys. Lett. B}\ }\textbf {\bibinfo {volume} {256}},\ \bibinfo {pages} {121}
  (\bibinfo {year} {1991})},\ \bibinfo {note} {[Erratum: Phys.Lett.B 262, 497
  (1991)]}\BibitemShut {NoStop}%
\bibitem [{\citenamefont {Abe}\ \emph {et~al.}(1992)\citenamefont {Abe} \emph
  {et~al.}}]{r38}%
  \BibitemOpen
  \bibfield  {author} {\bibinfo {author} {\bibfnamefont {F.}~\bibnamefont
  {Abe}} \emph {et~al.} (\bibinfo {collaboration} {CDF}),\ }\href {\doibase
  10.1103/PhysRevLett.69.3704} {\bibfield  {journal} {\bibinfo  {journal}
  {Phys. Rev. Lett.}\ }\textbf {\bibinfo {volume} {69}},\ \bibinfo {pages}
  {3704} (\bibinfo {year} {1992})}\BibitemShut {NoStop}%
\bibitem [{\citenamefont {Abachi}\ \emph {et~al.}(1996)\citenamefont {Abachi}
  \emph {et~al.}}]{r39}%
  \BibitemOpen
  \bibfield  {author} {\bibinfo {author} {\bibfnamefont {S.}~\bibnamefont
  {Abachi}} \emph {et~al.} (\bibinfo {collaboration} {D0}),\ }\href {\doibase
  10.1016/0370-2693(96)00067-6} {\bibfield  {journal} {\bibinfo  {journal}
  {Phys. Lett. B}\ }\textbf {\bibinfo {volume} {370}},\ \bibinfo {pages} {239}
  (\bibinfo {year} {1996})}\BibitemShut {NoStop}%
\bibitem [{\citenamefont {Wu}\ \emph {et~al.}(2019)\citenamefont {Wu},
  \citenamefont {Liu}, \citenamefont {Geng}, \citenamefont {Hiyama},\ and\
  \citenamefont {Valderrama}}]{r42}%
  \BibitemOpen
  \bibfield  {author} {\bibinfo {author} {\bibfnamefont {T.-W.}\ \bibnamefont
  {Wu}}, \bibinfo {author} {\bibfnamefont {M.-Z.}\ \bibnamefont {Liu}},
  \bibinfo {author} {\bibfnamefont {L.-S.}\ \bibnamefont {Geng}}, \bibinfo
  {author} {\bibfnamefont {E.}~\bibnamefont {Hiyama}}, \ and\ \bibinfo {author}
  {\bibfnamefont {M.~P.}\ \bibnamefont {Valderrama}},\ }\href {\doibase
  10.1103/PhysRevD.100.034029} {\bibfield  {journal} {\bibinfo  {journal}
  {Phys. Rev. D}\ }\textbf {\bibinfo {volume} {100}},\ \bibinfo {pages}
  {034029} (\bibinfo {year} {2019})},\ \Eprint
  {http://arxiv.org/abs/1906.11995} {arXiv:1906.11995 [hep-ph]} \BibitemShut
  {NoStop}%
\bibitem [{\citenamefont {Guo}\ \emph {et~al.}(2014)\citenamefont {Guo},
  \citenamefont {Mei\ss{}ner},\ and\ \citenamefont {Wang}}]{r43}%
  \BibitemOpen
  \bibfield  {author} {\bibinfo {author} {\bibfnamefont {F.-K.}\ \bibnamefont
  {Guo}}, \bibinfo {author} {\bibfnamefont {U.-G.}\ \bibnamefont
  {Mei\ss{}ner}}, \ and\ \bibinfo {author} {\bibfnamefont {W.}~\bibnamefont
  {Wang}},\ }\href {\doibase 10.1088/0253-6102/61/3/14} {\bibfield  {journal}
  {\bibinfo  {journal} {Commun. Theor. Phys.}\ }\textbf {\bibinfo {volume}
  {61}},\ \bibinfo {pages} {354} (\bibinfo {year} {2014})},\ \Eprint
  {http://arxiv.org/abs/1308.0193} {arXiv:1308.0193 [hep-ph]} \BibitemShut
  {NoStop}%
\bibitem [{\citenamefont {Zyla}\ \emph {et~al.}(2020)\citenamefont {Zyla} \emph
  {et~al.}}]{r44}%
  \BibitemOpen
  \bibfield  {author} {\bibinfo {author} {\bibfnamefont {P.}~\bibnamefont
  {Zyla}} \emph {et~al.} (\bibinfo {collaboration} {Particle Data Group}),\
  }\href {\doibase 10.1093/ptep/ptaa104} {\bibfield  {journal} {\bibinfo
  {journal} {PTEP}\ }\textbf {\bibinfo {volume} {2020}},\ \bibinfo {pages}
  {083C01} (\bibinfo {year} {2020})}\BibitemShut {NoStop}%
\bibitem [{\citenamefont {Ablikim}\ \emph {et~al.}(2019)\citenamefont {Ablikim}
  \emph {et~al.}}]{r45}%
  \BibitemOpen
  \bibfield  {author} {\bibinfo {author} {\bibfnamefont {M.}~\bibnamefont
  {Ablikim}} \emph {et~al.} (\bibinfo {collaboration} {BESIII}),\ }\href
  {\doibase 10.1103/PhysRevD.100.111102} {\bibfield  {journal} {\bibinfo
  {journal} {Phys. Rev. D}\ }\textbf {\bibinfo {volume} {100}},\ \bibinfo
  {pages} {111102} (\bibinfo {year} {2019})},\ \Eprint
  {http://arxiv.org/abs/1906.00831} {arXiv:1906.00831 [hep-ex]} \BibitemShut
  {NoStop}%
\bibitem [{\citenamefont {Ablikim}\ \emph
  {et~al.}(2014{\natexlab{b}})\citenamefont {Ablikim} \emph {et~al.}}]{r46}%
  \BibitemOpen
  \bibfield  {author} {\bibinfo {author} {\bibfnamefont {M.}~\bibnamefont
  {Ablikim}} \emph {et~al.} (\bibinfo {collaboration} {BESIII}),\ }\href
  {\doibase 10.1103/PhysRevLett.112.022001} {\bibfield  {journal} {\bibinfo
  {journal} {Phys. Rev. Lett.}\ }\textbf {\bibinfo {volume} {112}},\ \bibinfo
  {pages} {022001} (\bibinfo {year} {2014}{\natexlab{b}})},\ \Eprint
  {http://arxiv.org/abs/1310.1163} {arXiv:1310.1163 [hep-ex]} \BibitemShut
  {NoStop}%
\end{thebibliography}%
\end{document}